\documentclass{emulateapj}
\providecommand{\mb}{$\Delta m_{15}(B)$}

\begin{document}
%%%%%%%%%% Title & Author %%%%%%%%%%%%%%%%%%%%%%%%%%%%%%%%%%%%%%%%%%%%%%%%%%%%%%%
\title{Constraints on Type I\lowercase{a} Supernova Progenitor Companions\\
 from Early Ultraviolet Observations with {\sl Swift} }
\author{Peter~J.~Brown\altaffilmark{1}, Kyle S. Dawson\altaffilmark{1}, 
David W. Harris\altaffilmark{1}, Matthew Olmstead\altaffilmark{1}, \\
Peter Milne\altaffilmark{2}, \& Peter W. A. Roming\altaffilmark{3}}
\altaffilmark{1} 
\altaffiltext{1}{Department of Physics \& Astronomy, 
University of Utah, 115 South 1400 East \#201, Salt Lake City, UT 84112, USA}            
\altaffiltext{2}{Steward Observatory, University of Arizona, Tucson, AZ 85719, USA}            
\altaffiltext{3}{Southwest Research Institute, Department of Space Science, 
San Antonio, TX 78238, USA}            
%\email{pbrown@physics.utah.edu}  

 %%%%%%%%%% Abstract %%%%%%%%%%%%%%%%%%%%%%%%%%%%%%%%%%%%%%%%%%%%%%%%%%%%%%%%%%%%
\begin{abstract}

We compare early ultraviolet (UV) observations of Type Ia Supernovae (SNe Ia) 
with theoretical predictions for the brightness of the shock associated with 
the collision between SN ejecta and a companion star.
Our simple method is independent of the intrinsic flux from the SN and
treats the flux observed with the  {\sl Swift}/Ultra-Violet Optical 
Telescope (UVOT) as conservative upper limits on the shock brightness. 
Comparing this limit with the predicted flux for various shock models, we constrain the geometry of the 
SN progenitor-companion system.  We find the model of a 1~M$_\sun$ red supergiant 
companion in Roche lobe overflow to be excluded at a 95\% confidence level for 
most individual SNe for all but the most unfavorable viewing angles.  
For the sample of 12 SNe taken together, the upper limits on the viewing angle 
are inconsistent with the expected distribution of viewing angles for RG stars 
as the majority of companions with high confidence.   The separation distance 
constraints do allow MS companions.  A better understanding of the UV flux 
arising from the SN itself as well as continued UV observations of young 
SNe Ia will further constrain the possible progenitors of SNe Ia.
\end{abstract}
%%%%%%%%%% Keywords %%%%%%%%%%%%%%%%%%%%%%%%%%%%%%%%%%%%%%%%%%%%%%%%%%%%%%%%%%%%
%%%%%%%%%%%%%%%%%%%%%%%%%%%%%%%%%%%%%%%%%%%%%%%%%%%%%%%%%%%%%%%%%%%%%%%%%%%%%%%%%
%\clearpage

\section{Type I\lowercase{a} Supernovae and their Progenitor Systems}  \label{intro}
 
From the first detections of the acceleration of an expanding universe \citep{Riess_etal_1998,Perlmutter_etal_1999}, 
Type Ia Supernovae (SNe Ia) have continued to be the best 
probes of the distant universe for measuring cosmological parameters 
(see recent results in e.g. \citealp{Riess_etal_2011,Sullivan_etal_2011,Suzuki_etal_2011}).  
They are useful as standardizable candles because of the well established 
empirical relationship between the absolute brightness and other observables 
such as the light curve shape and colors (cf. \citealp{Phillips_etal_1999,Jha_etal_2007,  Guy_etal_2007}).  
Hundreds of SNe later, SN cosmology is now limited by systematic rather than 
statistical errors \citep{Wood-Vasey_etal_2007,Kessler_etal_2009, Conley_etal_2011}.

One systematic error for using SNe Ia at cosmological distances may arise from redshift evolution  
in the SN explosions due to differences 
in the average properties (mass, metallicity, etc.) of the progenitor systems \citep{Mannucci_etal_2006,Howell_etal_2007}.
This is especially worrisome because 
the underlying progenitor system for a SN Ia explosion is still unknown, 
with two favored progenitor scenarios \citep{Livio_1999}.  
In the single degenerate case, a carbon-oxygen white dwarf (WD) accretes material 
from a main sequence (MS) or red giant (RG) companion star and explodes when it 
nears the Chandrasekhar mass \citep{Whelan_Iben_1973,Nomoto_1982}.
Alternatively, the double degenerate scenario involves two WDs which merge and 
explode \citep{Webbink_1984,Iben_Tutukov_1984}.
Determining the nature of the progenitor systems of SNe Ia is 
critical to confidently and precisely use them as cosmological standard candles.

In an effort to improve our understanding of the progenitors in the single degenerate scenario, 
several groups have studied various effects of a SN Ia explosion on a companion star.
These include the amount of hydrogen that is stripped from the companion and 
possibly detectable in observations of the SN spectrum 
(see e.g. \citealp{Leonard_2007,Marietta_etal_2000}) 
or observable effects on the leftover companion such as metallicity differences 
or an abnormal velocity \citep{Meng_etal_2007}.  
Approaching the problem from a different angle, Kasen (2010; hereafter K10) 
explores the SN-companion shock itself and the resulting radiation that such an 
interaction might create for various progenitor systems.   
Shown to be similar in timescale and luminosity to the shock breakout of core-collapse SNe in K10, 
this shock refers to the SN shock wave impacting the surface of the companion star.
The K10 predictions provide an alternative route to learning about the elusive companions.

The K10 models inspired several groups to look for evidence of shock emission in their existing data sets.
 \citet{Hayden_etal_2010a} analyzed nearly 500 SNe Ia with rest-frame $B$-band observations from the 
 Sloan Digital Sky Survey II SN survey \citep{Frieman_etal_2008}.  
They compared  the observed optical flux 
with simulations to show that RG stars are disfavored as the dominant companion.
Rather, the majority of systems must have MS companions of less than 6~M$_\sun$ 
and/or a second WD (the double degenerate scenario).
Similarly, \citet{Tucker_etal_2011} analyzed 695 light curves of low and high redshift SNe Ia 
from a variety of sources in the rest frame UBVRI and found no evidence for shock emission from RG companions. 
\citet{Mo_etal_2011} also analyzed early light curves and saw no shock emission, but also recognized a 
strong degeneracy between the constraints and the assumed shape of the early unshocked light curves.

The high luminosity of the shock emission in the ultraviolet (UV) predicted from RG companions in this 
model should be easily seen in nearby SNe.
 Despite the smaller sample size and our limited understanding of SN light curves in the UV, 
we show that early UV observations can put strong constraints on the progenitor systems of SNe Ia.
By explicitly accounting for the effect of viewing angle on luminosity, we are able to constrain 
the companion systems of individual SNe rather than just statistical constraints on the sample as a whole.  
This paper presents these constraints on UV shock emission as follows:
in Section \ref{model} we describe our use of the K10 model.
Section \ref{analysis} describes the {\sl Swift}/Ultra-Violet/Optical Telescope (UVOT) 
observations \citep{Gehrels_etal_2004,Roming_etal_2005}, the determination of the date of explosion, and the method to constrain the SN Ia companion 
system necessary to explain the observed UV flux.
Our discussion of the progenitor constraints and how they could be restricted further 
is in Section \ref{discussion}.  
%%%%%%%%%%%%%%%%%%%%%%%%%%%%%%%%%%%%%%%%%%%%%%%%%%% 
 %%%%%%%%%%%%%%%%%%%%%%%%%%%%%%%%%%%%%%%%%%%%%%%%%%%%%%%%%%%%%%%%%%%%%%%%%%%%%%%%% 
 
\section{Theoretical Predictions for the Brightness of the Shock} \label{model}

%%%%%%%%%%%%%%%%%%%%%%%%%%%%%%%%%%%%%%%%%%%%%%%%%%%%%%%%%%%%%%%%%%%%%%%%%%%%%%%%% 

In the K10 model, the SN ejecta interact with the companion star that fills its Roche lobe.  
This shock heats the companion's surface which faces the explosion.  
As the SN ejecta continue to expand past the companion, the presence of the companion results in a cone shaped hole 
in the ejecta from which radiation from the shocked material escapes.  
This results in a prompt X-ray burst and a continued diffusion of thermal energy emitted at UV/optical wavelengths.  
K10's analytic derivation of the evolution of the luminosity (L) and temperature (T) 
in days after the explosion (t) depend primarily on the separation distance 
a$_{13}$ ($a/10^{13} cm$) between the SN progenitor and its companion.
These equations, 22 and 25 from K10, are reproduced below.  
%\begin{equation}\label{...} ... \end{equation}

\begin{equation}\label{eq_L}L_{c,iso} = 10^{43} a_{13} M_c^{1/4} v_9^{7/4} t_{day}^{-1/2} {\rm ergs~s}^{-1}  \end{equation}

\begin{equation}\label{eq_T}T_{eff} = 2.5 \times 10^4 a_{13}^{1/4} \kappa_e^{-35/36} t_{day}^{-37/72} \end{equation}

In the equations above, the SN ejecta mass  $M_c$ (in units of the Chandrasekhr mass), 
the SN expansion velocity~$v_9$ (in units of $10^9$ cm/s), 
and the electron scattering opacity  $\kappa_e$, are all assumed to be 1 for normal SNe Ia. 
 Because the companion is assumed to be filling its Roche lobe, its radius and approximate initial mass 
can be determined from the separation distance.
% {\bf explain more} 

K10 also calculated a series of numerical spectra and light curves with a 3d radiation transfer code.  
This is done for three models: a 1~M$_\sun$ red giant with a$_{13}$=2, a 6~M$_\sun$ 
main sequence star with a$_{13}$=0.2, and a 2~M$_\sun$ main sequence star with a$_{13}$=0.05.  
These numerical models include the emission from the shock as well as the SN. 
Most importantly for our analysis, these models also capture the asymmetry of the emission, 
or equivalently, the orientation of the SN and the companion with respect to the observer. 
The peak brightness occurs for a viewing angle of 0 degrees, 
where the companion lies directly along the line of sight between the observer and the SN explosion.  
The light curves displayed in K10 show the increase in luminosity for larger companions 
(at a larger separation) and shorter wavelength observations.  

We begin with these numerical models of K10 and perform spectrophotometry 
on the spectra scaled to 10 parsecs to yield absolute magnitudes in the UVOT bands.  
A comparison with observed UV light curves of SNe Ia \citep{Brown_etal_2009, Milne_etal_2010} 
shows that the SN component of the model peaks brighter 
than most observed SNe, which show great diversity.  
This discrepancy is likely due to incomplete line lists for the iron-peak elements whose absorption blanket the UV (Kasen 2011, private communication). 
A detailed comparison between theoretical models and the growing sample of UV photometry and spectroscopy \citep{Brown_etal_2009,Milne_etal_2010,Bufano_etal_2009,Foley_etal_2011} is beyond the scope of this paper. 

\begin{figure} 
%\resizebox{16cm}{!}{\rotatebox{90}{\includegraphics*{f2.eps}  } }
\plottwo{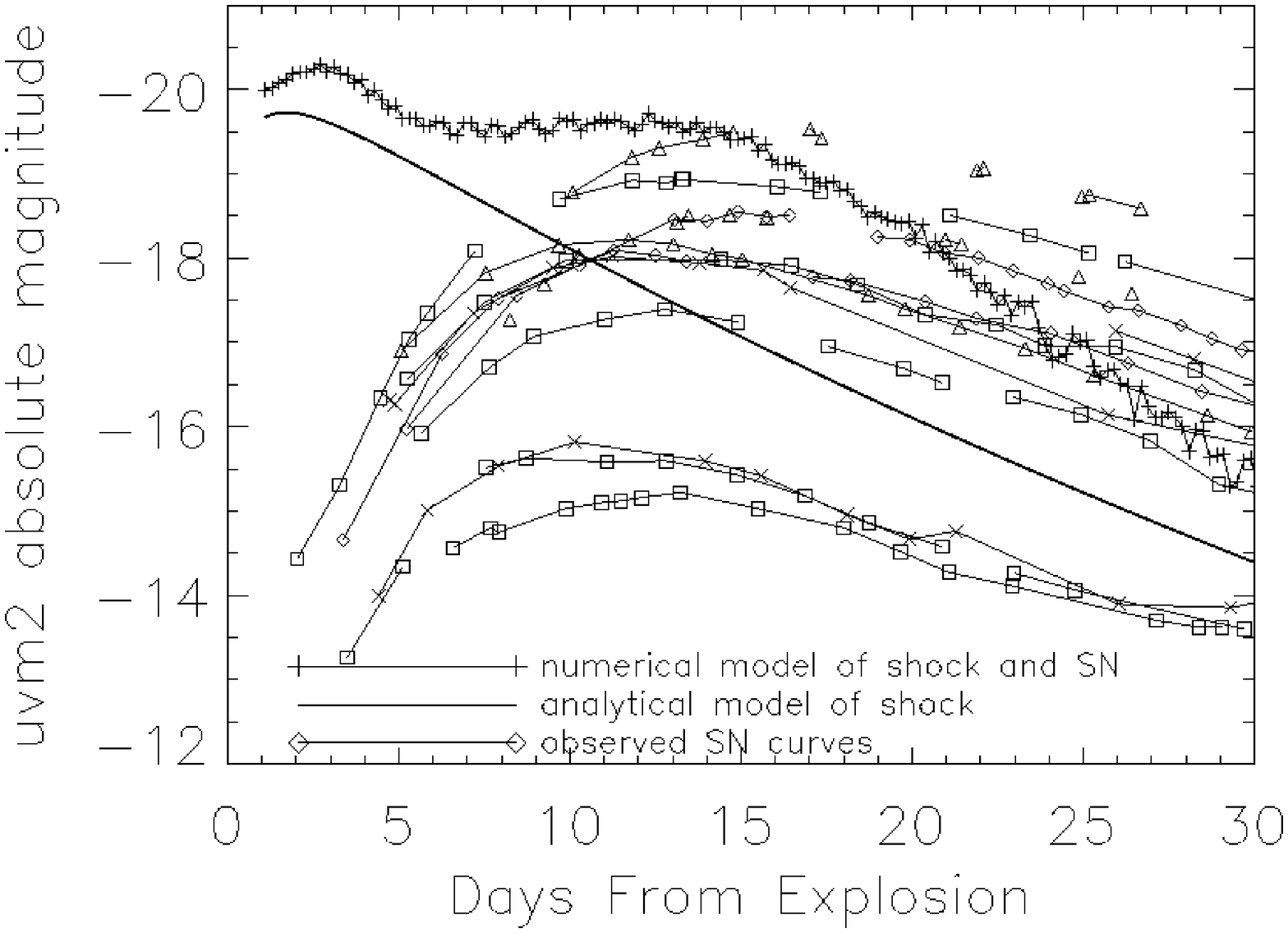}{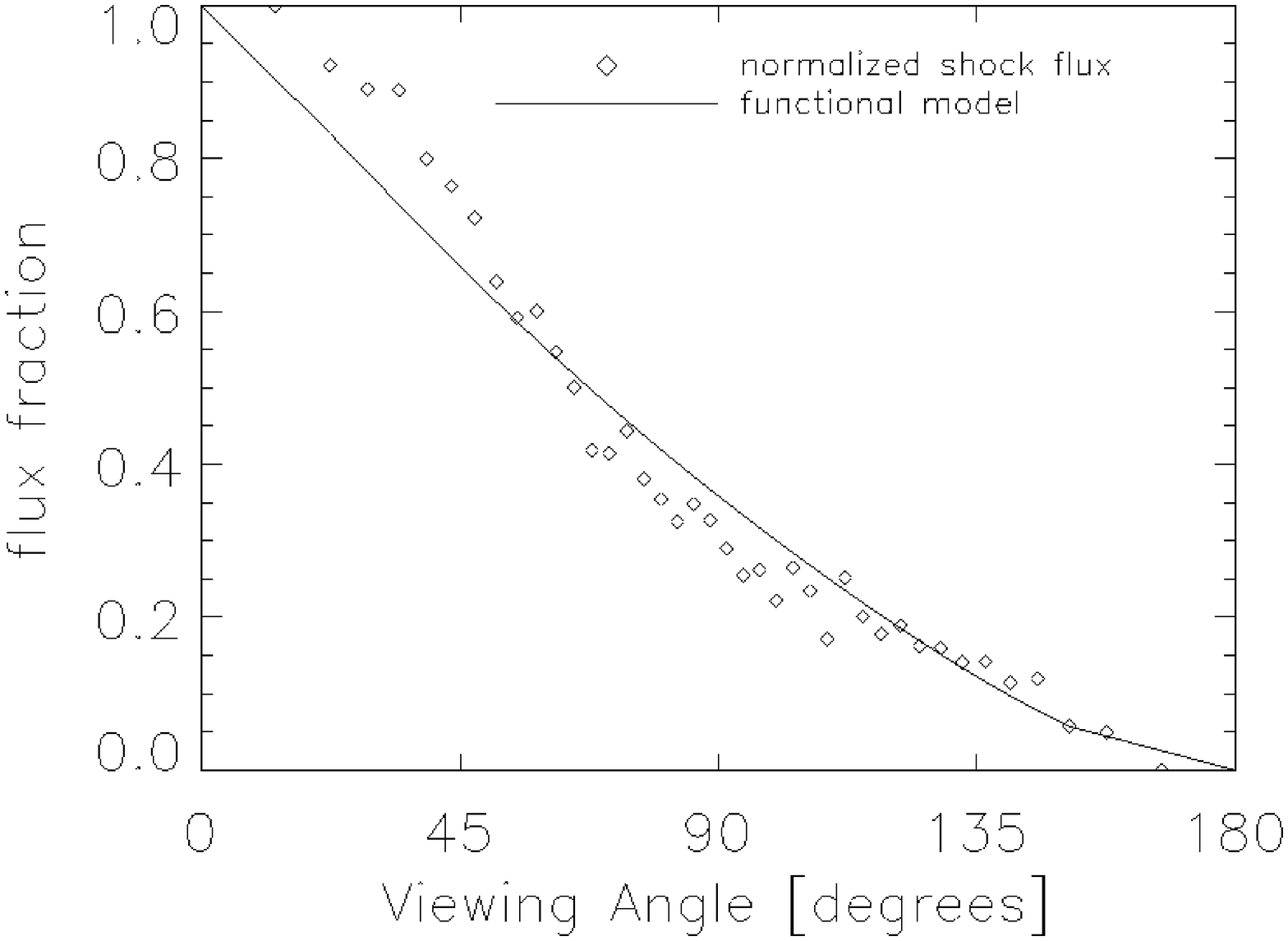} 
\caption[Results]
{  {\it Left:} Numerical simulation (including the shock and subsequent SN light curve at a viewing angle of 13 degrees) 
and the analytic model of the shock (both from K10) in the uvm2 filter.     
The luminosity of the analytical and numerical models is similar at early times when the shock luminosity dominates.  
The absolute uvm2 magnitudes for our sample of SNe is also shown.
As observed, the UV light curves actually peak significantly fainter than predicted in the models 
and show great diversity. 
{\it Right:} Ratio of the shock flux seen at each viewing angle to the shock flux seen nearly straight on (looking down on the companion) 
is shown with open diamonds.  
The formula used in the following analysis is shown with the solid line.  
The divergence at small angles is because our function is forced to be unity at 0 degrees and zero at 180 degrees. } \label{plot_lc_comparison}
\end{figure}

To reproduce the brightness of the shock, we use equations 1 and 2 
to create a temporal 
series of blackbody spectra with the appropriate temperature and luminosity 
for various values of $a_{13}$. 
In the left panel of Figure \ref{plot_lc_comparison} we show  
the uvm2 light curve of the $a_{13}=2$  numerical model (viewed nearly face-on) along with the analytic model and 
absolute magnitudes from our observed sample (see below).  
The fainter, later 
rise of the SN in the UV means that the contrast between the shock and the SN should be much 
stronger, later, and at fainter magnitudes than expected from the numerical models shown in Figure 3 of K10. 
This contrast allows us to put tight constraints on the shock emission even with a small sample.

%%%%%%%%%%%%%%%%%%%%%%%%%%%%%%%%%%%%%%%%%%%%%%%%%%%%%%%%%%%%%%%%%%%%%%%%%%%%%%%

Because our sample of SNe observed in the UV is small, and the contrast between shock and SN luminosity so high, 
we determine the dependence 
of the luminosity on viewing angle explicitly and can thus put constraints on individual SNe.  
This is in contrast to other analyses \citep{Hayden_etal_2010b,Bianco_etal_2011,Tucker_etal_2011} 
which statistically addressed the $\sim10\% $ observable fraction for which the shock would be brightest.  
To estimate the dependence of the luminosity on the viewing angle, we use the luminosity in the UV range (2000-4000 \AA) 
between 2-4 days after explosion from the smallest of the K10 numerical models for which a$_{13}=0.05$ 
for a range of viewing angles.
At longer wavelengths and later epochs, the SN+shock flux is dominated by the SN flux.  
We first isolate the shock flux by subtracting the SN dominated flux 
(from the largest off axis angle) from the total flux at each angle. 
This is normalized by the shock flux viewed at the optimal viewing angle.
We find the ratio to be roughly proportional 
to the cosine of the viewing angle (in radians) multiplied by an additional damping parameter:

\begin{equation}\label{eq_theta} f = (0.5 cos~\theta + 0.5) \times ( 0.14~\theta^2 -0.4~\theta + 1). \end{equation}

The fractional flux values (f) from the model and the fit function are displayed 
in the right panel of Figure \ref{plot_lc_comparison}.  
While the relation is not physically motivated, it serves as an approximate fit to the data 
and a means to incorporate the viewing angle dependence into the analytic expressions 
for the shock brightness.
The ratios for the models with larger separation fall off more slowly with viewing angle, thus making our use 
of this function more conservative for the viewing angles constrained below.
We use this angular dependence from the numerical models with the analytic expressions for temperature and luminosity 
to produce the modeled blackbody spectrum of the shock for a given epoch.
This model is then compared to the UVOT data to constrain the separation distance and viewing angle 
allowed in the single degenerate, Roche-lobe scenario for each SN Ia.

 %%%%%%%%%%%%%%%%%%%%%%%%%%%%%%%%%%%%%%%%%%%%%%%%%%%%%%%%%%%%%%%%%%%%%%%%%%%%%%%%% 
 
\section{Model Constraints from Observations}\label{analysis}

For this study, we use UV and optical observations of 12 nearby (z $<$ 0.03), 
spectroscopically classified SNe Ia obtained with the Swift/UVOT.  
These SNe were selected from the full sample of template subtracted SNe Ia 
(as of Apr 2011) based on having 
UVOT observations within ten days of the estimated time of explosion. 
The photometry for seven of these SNe has been previously published in \citet{Brown_etal_2009} 
and \citet{Milne_etal_2010}, according to the UVOT calibration given in \citet{Poole_etal_2008}.  
We also present UVOT photometry for five newer SNe in Table 1, 
reduced using the method outlined in \citet{Brown_etal_2009}, 
including subtraction of the host galaxy flux.  
To aid in the determination of the time of explosion, we have 
added ground based $B$-band observations from \citet{Pastorello_etal_2007}
and \citet{Stritzinger_etal_2010} to the UVOT observations of SNe 2005cf and 2006dd, respectively.  
Characteristics of all SNe in our study, including \mb, the peak B band absolute magnitude, the host galaxy identification, 
and host galaxy morphology, are listed in Table 2.

%Near-peak observations of SN~2007gi from \citet{Zhang_etal_2010} are also included to better 
%determine the peak time of this SN, as the UVOT observations ended before then. 
We use distance and extinction 
estimates previously used to study the absolute magnitudes 
at maximum light \citep{Brown_etal_2010} to determine the absolute magnitudes at 
each epoch of observation.  Values for the newer SNe were calculated in the same manner.  
When available, distances calculated from surface brightness fluctuations (SBF) in the host 
galaxy are used.  In the absence of such distance measures, the host galaxy recessional velocity, 
corrected for the local velocity field \citep{Mould_etal_2000}, 
is converted to a distance using a value of H$_0$=72 km/s/Mpc \citep{Freedman_etal_2001}. 
The uncertainty in the Hubble flow distance includes the stated error 
in the corrected velocity (on the order of 30 km sec$^{-1}$) and a typical peculiar motion of 150 km sec$^{-1}$.
The reddening is determined by the peak colors \citep{Phillips_etal_1999}.  
The extinction is corrected using the Milky Way extinction law from \citet{Cardelli_etal_1989} 
using the coefficients appropriate for a SN Ia and the UVOT filters given in \citet{Brown_etal_2010}.
The distance and extinction values used for the SNe are given in Table 2. 

These UV and optical data are used to determine the explosion dates and constrain the possible 
separation distances and the viewing angle of the progenitor systems.
Though UV grism spectroscopy is available for some of these SNe \citep{Bufano_etal_2009,Foley_etal_2011}, 
the sample is much smaller, especially at the early times needed to see evidence of the shock.  
Furthermore, the thermal shock is expected to exhibit a mostly broad-band effect, 
so we limit the analysis to the UVOT photometry.

\subsection{Determining the explosion date}\label{explosiondate}

Since the luminosity of the shock rises and falls quickly compared to the SN light, we require an accurate determination of the explosion date.  
%For now: The time of explosion is estimated to be 16.82 days before the peak in the B band light curve.  
%A one $\sigma$ uncertainty of 1.77 days is used to allow for the distribution of 
%rise times for SNe Ia from \citep{Hayden_etal_2010a}, which is also consistent with \citet{Garg_etal_2007} and the global mean of \citet{Strovink_2007}.  
%This will be replaced with:
To estimate the time of explosion, we adopt a method similar to \citet{Hayden_etal_2010b}. 
The MLCS2k2 \citep{Jha_etal_2007} $B$-band template is extrapolated to a date of explosion 16.5 days before peak by assuming the flux rises from zero proportionally to the square of the time since explosion.  
This template is fitted to the early UVOT $b$-band or ground based $B$-band data with the peak flux, time of peak flux, and a multiplicative stretch factor as free parameters.   
We use epochs from the time of explosion to 5 days after maximum light in order to constrain the time of peak flux
without biasing the stretch of the template used to estimate the time of explosion by data taken well after maximum light.  

The accepted parameters are those resulting in the lowest $\chi^2$ between the template and the observations.
%The $\chi^2$ values for each set of variables is translated into a probability for the corresponding explosion date.  
After an initial fit to center the parameter grid, we perform a fit using the same procedure 
on 1000 Monte Carlo realizations of the optical data with the associated errors.  
%The template is also stretched to match the average rise time and scatter so that SNe with little data on the rising branch will receive a value equal to the average rise time and an error equal to the scatter.
This results in an array of possible explosion dates from which we draw during the Monte Carlo simulations described below.  
The mode of the dates (binned to 0.1 days) and the 95\% bounds are given in Table 3.  
For our sample, the mean rise time is 15.50 days with a standard deviation of 1.73 days, a day shorter than the 16.82 day rise time and scatter of 1.77 days \citet{Hayden_etal_2010b} found for a larger, low z sample.

\subsection{Comparing the observations to the models}\label{comparisons}

As shown in Section \ref{model}, numerical models do not accurately reproduce the light curves seen in UV observations of SNe Ia.  \citet{Foley_etal_2011} also found discrepancies between the early UV light curves of SN~2009ig and 
the fireball model that were not consistent with shock interaction.
In addition, the current UV templates \citep{Milne_etal_2010} are made from many of the same SNe  
used in the current analysis.  
We therefore take a conservative course and use the observed UVOT flux 
as an absolute upper limit on the emission from the shock.
The absolute magnitude measured at each epoch past explosion is compared to the shock model from 
Section \ref{model} to constrain the viewing angle for each separation distance, a$_{13}$.
To determine the confidence intervals of the constraints, we perform 2000 Monte Carlo realizations of each SN light curve, 
varying the absolute magnitude for each epoch with the photometric error, extinction error, and the uncertainty in the distance modulus, 
and varying the days after explosion with the uncertainty of the explosion date.  

For each combination of a$_{13}$ and $\theta$, we count the number of Monte Carlo realizations 
that produce a flux that is less than the modeled shock flux.  
This is displayed graphically in Figure 2 where we compare the observed absolute magnitudes 
with various models at a fixed angle (left panel) and with various angles for a fixed model (right panel).  For each value of a$_{13}$, the value of $\theta$ for which 95\% of the realizations exclude the given model is considered the 95\% exclusion limit.  This is done for each epoch and filter.  The strictest results map the area excluded at 95\% confidence in a$_{13}$-$\theta$ parameter space for each SN.

For typical observation lengths, the first epoch uvm2 observations are the most 
constraining because the shock is brighter and the SN fainter for shorter wavelengths 
and earlier epochs.  
The constraints from the uvw2 filter, though centered at a shorter wavelength than the uvm2 filter, are 
typically weaker because the red tails of the filter \citep{Brown_etal_2010,Breeveld_etal_2011} 
allow significant optical light from the rising SN. 
Because the SN flux is included in our upper limits, the brightness of the SN rather 
than the depth of the observation is usually the limiting factor.
Subtracting the SN flux 
would allow stronger constraints from the uvw2 without any special treatment for the red tail of the filter.
%The earliest observations sometimes yield only upper limits in the UV, so observing 
%strategies weighting more time to the UV filters, in particular uvm2, have been 
%pursued for more recent SNe.

Because the shock emission fades quickly with time, the factor which dominates 
the constraints is how soon after explosion the first UV observation occurs.  
This can be seen in Figure 2 where the consequence of excluding earlier, fainter observations is clear.
For each day that the first observation is delayed, 
the allowed companion separation distance for a given angle approximately doubles.  
Clearly, discovering young SNe and announcing them quickly, coupled with a fast 
turnaround in observing in the UV, is important to driving these constraints further.

%%%%%%%%%%%%%%%%%%%%%%%%%%%%%%%%%%%%%%%%%%%%%%%%%%% 

The results for all of our SNe using the most constraining epoch and UVOT filter are shown in the left panel of Figure 3. 
In Table 3 we list 50 and 95\% lower limits on the viewing angle 
for two cases: a$_{13}$=2 (corresponding to the 1~M$_\sun$ RG case) and a$_{13}$=0.2 
(corresponding to a 6~M$_\sun$ MS companion).
For many individual SNe, the RG scenario is only allowed for the most unfavorable viewing angles.  
For example, the solid angle corresponding to viewing angles greater than 135 degrees covers only 10\% of a sphere, 
yet eleven of the twelve SNe studied here require viewing angles greater than that for the RG scenario.

In the right panel of Figure 3, we display cumulative distribution functions (CDFs) of the 95\% lower limits from the 
a$_{13}$=2 and 0.2 models compared to what would be expected for the sample of random viewing angles.  
The angles that would be expected from a random distribution of observations are not determined solely by geometry, 
but are also dependent on the observational errors and the determination of 95\% limits.
Within the Monte Carlo simulation, we apply the same uncertainties in reddening, distance, and explosion date to the 
a$_{13}$=2 and 0.2 models at each angle.  We then compare the flux with that from the nominal model (viewing angle equal to zero) and use the viewing angle 
sensitivity function to compute the apparent angle.  Thus for each input angle we can map 
its random probability to a 95\% lower limit on the angle.
A noiseless measurement would produce a median value of 90 degrees 
(similar to the 50\% curve for which points scatter equally above and below the line), 
but the consideration of errors and 
determining 95\% exclusion regions pushes the curve to the left, 
allowing more instances of small viewing angles.  
The 95\% limits from random angles for the MS case 
shift a little further to the left as the photometric errors are a larger fraction of the model flux. 
The shift is small, however, because the uncertainty is dominated by the extinction errors.

Quantitatively comparing the CDFs is non-trivial.  The 95\% lower limits from the models are for the shock emission, 
while the limits from the observations are from the shock plus SN emission.  Thus the true angular distribution should 
lie to the right of the observational limits.  A Kolmogorov-Smirnov (K-S) test on these CDFs 
gives upper limits to the probability that 
they arise from the same distribution, but are therefore only valid when the observed CDF is to the right of the predicted CDF.  
The a$_{13}$=2 RG case is clearly excluded, with a maximum separation between the observed limits 
and the expected distribution of D=0.888  and a negligible probability that they come from the same distribution.  
Testing models with progressively smaller separation distances, we find that models 
with separation distances a$_{13}>$0.4 are excluded at a 95\% confidence level.
For the  a$_{13}$=0.2 case, the observed lower limits on the viewing angle drop below that expected from the 
random angle distribution, so we are unable to place constraints without subtracting the SN light.  

These constraints for the whole sample assume a single species of companion stars.  
RG companions could account for a fraction of the systems, while MS stars or white dwarfs account for the 
majority of the companions.  
Having observed no possible RG companions with a viewing angle less than 90 degrees 
(encompassing 50\% of viewing angles but 80\% of the lower limits after accounting for the observational errors) 
in our sample of 12 SNe, 
we use Poisson statistics to constrain the fraction of RG companions to less than 31\% of systems at 95\% probability.

To address the possibility that some classes of SNe Ia may result from different progenitor systems, we have repeated the above analysis with 
the 8 SNe with \mb ~between 1.0 and 1.5.  This excludes the very broad SN~2009ig and the three rapid decliners. The results are nearly identical to that of the whole sample, with our new 95\% confidence limits excluding companions with a$_{13}>$0.5 (rather than 0.4).  Further differences in SNe may correspond to their host environment, as differences in the absolute magnitudes and light curve shape of SNe Ia appear to correlate with host galaxy mass and star formation history \citep{Lampeitl_etal_2010, Kelly_etal_2010,Sullivan_etal_2010}.  Our sample is not large enough to make conclusions on the progenitor systems of individual subsets, defined by host galaxy and SN characteristics. This may be possible with the larger optical samples \citep{Hayden_etal_2010b,Bianco_etal_2011,Mo_etal_2011}.   A larger UV sample would allow the above statistical tests on SNe Ia divided by photometric, spectroscopic, and host galaxy properties.  We reiterate that the strength of this analysis lies in its ability to place limits on the progenitor systems of individual SNe.  For our small sample, no SNe show the signatures expected from RG companions.  This includes normal SNe Ia from the full range of light curve widths, but not the subclasses of SN~1991T-like, 2000cx-like, 2002cx-like, or probable super-Chandrasekhar mass SNe.

\subsection{Looking for Evidence of Shocks in UV-Optical Colors }\label{explosiondate}

The presence of shock emission would also be detectable by a distinct change in the colors at early times.  
The shock would initially be quite blue, and then redden as it fades.  The intrinsic light of the SN, on the other hand, 
begins quite red and becomes bluest just before the SN luminosity peaks.  Since the colors of the numerical models do not match the observations, we test for color evolution from the shock by adding the shock flux of various models (the RG case for multiple viewing angles and the two MS cases at the optimum viewing angle) to the observed flux of SN~2009ig, the SN in our sample with the earliest observations.  The resulting uvw1-v color curves (from the SN+shock flux) are displayed in the left panel of Figure \ref{plot_colors}.  The observed colors of our sample are added to those in the right panel. The observed colors do not show the early blue colors of the RG companion case at most viewing angles.  The a$_{13}=$2 model at 135 degrees and the a$_{13}>$0.2 model show a local (red) maximum in the color that is not seen in any of the observed color curves, though only half of them begin early enough to see such a feature.  Qualitatively the colors appear consistent with our conclusions above, namely that systems with RG companions would have to be viewed from statistically improbable angles.  Because of the apparent intrinsic diversity in the colors, however, it is not currently possible to place quantitative limits on the shock for individual SNe.  The colors could provide useful evidence to support or refute a possible shock seen in a light curve.  If the intrinsic colors are better understood through modeling or finding a SN with otherwise similar properties (but no suggestion of a shock at early times), the degeneracy between the companion separation and the viewing angle might be broken through a comparison of the colors. 

\section{Conclusions and Future Work} \label{discussion}

Using Swift/UVOT observations of SNe Ia taken less than 10 days after explosion, 
we have placed new constraints on the companion in the Roche lobe overflow, single degenerate scenario.
We used the numerical models of K10 
coupled with the analytic models of K10 to predict the light curves of 
UV shock emission as a function of separation distance and viewing angle. 
For all individual SNe Ia with early observations, we are able to constrain 
the viewing angle to be greater than 112 degrees at 95\% confidence for separation distances a$_{13}>2$.  
For most of the SNe, the lower limit on the viewing angle is greater than 160 degrees.
Comparing the distribution of 95\% constraints from the full sample of 12 SNe to a 
distribution expected from a random sample of viewing angles, 
we exclude the model of a red giant companion in Roche lobe overflow with extremely high confidence.  
Our limits allow the companion to be at a separation distance less than a$_{13}>0.4$ for the whole sample
and less than a$_{13}>0.5$ for the SNe Ia with normal light curve widths.
These limits are comparable to those of \citet{Hayden_etal_2010b} and 
\citet{Bianco_etal_2011}, but without any assumptions on the intrinsic SN flux.
Additionally, because we explicitly account for the angular dependence of 
the flux, we can constrain not only the progenitor separation for the sample as a whole, but for 
individual SNe.  

An excellent example of the progenitor system constraints that can be determined for individual SNe this way is the recently discovered SN~2011fe, 
for which pre-explosion HST imaging and very early UV, optical, and X-ray observations give very tight constraints 
on progenitor systems \citep{Li_etal_2011, Nugent_etal_2011, Horesh_etal_2011}.  
Using the same technique as above on {\sl Swift}/UVOT observations of SN~2011fe about one day after explosion, we rule out even solar mass main sequence companions in the Roche-lobe limit scenario 
with a separation distance constraint of a$_{13}<0.01$\citep{Brown_etal_2011}.  Optical observations a mere four hours after the estimated explosion date constrain it by a factor of ten further \citep{Bloom_etal_2011}.

\citet{Sternberg_etal_2011} recently reported a preference for SNe Ia to have blue shifted sodium absorption lines.  
Because of their velocities, this absorption is attributed to shells ejected during nova explosions that would 
occur from mass accretion in the single degenerate scenario.  Thus at least some SNe Ia likely occur in systems 
with non-degenerate companions.  As shown here, most of those companions must be MS stars.  Our study has several objects 
in common with the \citet{Sternberg_etal_2011} study: SNe 2007af, 2008ec, 2008hv, 2009ig, and SNF20080514-002.
Two of these exhibit blue shifted lines; however, several of the SNe are expected to have 
blue shifted lines from intervening clouds of gas with random directions just as several have redshift lines.
It is the preference for blue shifted lines in the sample that leads to the conclusion, so statements cannot be 
made for individual SNe.  The observance of time variable absorption lines, as seen in several SNe 
(e.g. \citealp{Patat_etal_2007}) for SNe with early UV observations would be able to constrain the companions 
from different directions.

The limits presented here could be improved if the underlying SN light in the UV, 
including extinction, were better understood.  
Subtracting the SN light would put stricter limits on the flux from interaction with MS companions.  
Developing UV SN templates is complicated by the fact that the UV flux is strongly affected by line blanketing, 
so small differences in composition and density can have a drastic effect on the 
UV luminosity \citep{Lentz_etal_2000,Sauer_etal_2008}.  As shown here, even the 
average SN UV light curve is not easily reproduced by current modeling.  
The effect of extinction uncertainties could be reduced by better understanding the intrinsic UV colors.
Alternatively, if the peak UV luminosities are better understood, the shock luminosity 
could be compared to the peak SN luminosity which would be similarly affected by extinction.

The analytic model used here assumed a constant opacity from electron scattering (K10).  
The close correspondence of the early time UV light curves from the analytic model with 
those from the numerical models (which 
includes a more realistic line opacity) suggests this assumption is not far off.  
However, to improve the model for the shock,  a better 
understanding of the time(temperature)-dependent opacity in the interacting 
material is essential.
To test the sensitivity of our method to a decrease in the luminosity due to an increased opacity, 
we reduced the brightness of the shock flux by a factor of ten, 
approximately the factor by which the observed SN flux is reduced relative to the model.
The red giant case is still excluded with the KS test giving a maximum separation of 0.64 
and a probability of $4\times 10^{-5}$.  
Thus the main conclusion is still robust if 
the model flux is off by an order of magnitude.  

Finally, the brightness of the interaction has been considered for the model 
in which the progenitor companion fills its Roche lobe at the time of the SN Ia explosion.
\citet{Justham_2011} has suggested that the accretion from the companion could also 
deposit angular momentum onto the white dwarf.  This would result in rotational support 
and allow for a larger mass than a non-rotating white dwarf 
to undergo a SN Ia explosion.  If the mass transfer ends before the WD explodes as a SN, 
the time for the white dwarf 
to slow its rotation sufficiently to explode might allow the companion to decrease in size.  
Thus the companion would present a smaller target for the 
SN ejecta and produce a much smaller shock luminosity 
than the Roche lobe model considered here.  A more detailed analysis of the 
expected shock luminosity from an SN Ia explosion in such a system is required 
to compare it to observations.

While some of the SNe shown here were observed at very early epochs, most were 
observed by Swift following a discovery, confirmation, and reporting sequence 
that took several days.  
A more rapid dissemination of SN candidates by high cadence searches could yield 
a much larger sample of SNe to tighten the constraints on progenitor size.  
The Palomar Transient Factory (PTF) has proven to be highly efficient 
at finding young SNe Ia \citep{Cooke_etal_2011} and has already provided 
the youngest SN Ia ever discovered \citep{Nugent_etal_2011}.  
The rapid response and UV capability of Swift make it an excellent observatory 
for advancing the studies of UV shock emission in SN Ia explosions.
Future UV observatories should also 
consider short turn around target of opportunity programs to exploit the 
valuable information contained in the early discoveries of transient sources.

%%%%%%%%%%%%%%%%%%%%%%%%%%%%%%%%%%%%%%%%%%%%%%%%%%%%%%%%%%%%%%%%%%%%%%%%%%%%%%%%%
\acknowledgements
We are grateful to D. Kasen for providing his theoretical series of spectra and 
advising on the comparisons.  
This work at the University of Utah is supported by NASA grant NNX10AK43G, 
through the Swift Guest Investigator Program.
This analysis was made possible by access to the 
public data in the Swift data archive and the NASA/IPAC Extragalactic Database (NED). 
NED is operated by the Jet Propulsion Laboratory, California Institute of Technology, 
under contract with the National Aeronautics and Space Administration.  

%%%%%%%%%%%%%%%%%%%%%%%%%%%%%%%%%%%%%%%%%%%%%%%%%%%%%%%%%%%%%%%%%%%%%%%%%%%%%%%%%

%%%%%%%%%%%%%%%%%%%%%%%%%%%%%%%%%%%%%%%%%%%%%%%%%%%%%%%%%%%%%%%%%%%%%%%%%%%%%%%%%

%%%%%%%%%%%%%%%%%%%%%%%%%%%%%%%%%%%%%%%%%%%%%%%%%%%%%%%%%%%%%%%%%%%%%%%%%%%%%%%%%
%%%%%%%%%%%%%%%%%%%%%%%%%%%%%%%%%%%%%%%%%%%%%%%%%%
%  
%  %%%%%%%%%%%%      %%          %%%%%       %%         %%%%%%%%%%        %%%%
%      %%          %%  %%        %%  %%      %%         %%               %%     
%      %%         %%    %%       %%   %%     %%         %%              %%       
%      %%        %%      %%      %%   %%     %%         %%               %%      
%      %%       %%%%%%%%%%%%     %%%%%       %%         %%%%%%%           %%          
%      %%      %%          %%    %%  %%      %%         %%                 %%    
%      %%     %%            %%   %%   %%     %%         %%                  %%     
%      %%    %%              %%  %%  %%      %%         %%                 %%       
%      %%   %%                %% %%%%        %%%%%%%%%% %%%%%%%%%%       %%                      
%%%%%%%%%%%%%%%%%%%%%%%%%%%%%%%%%%%%%%%%%%%%%%
%%%%%%%%%%%%%%%%%%%%%%%%%%%%%%%%%%%%%%%%%%%%%%%%%%%%%%%%%%%%%%%%%%%%%%%%%%%%%%%
\begin{deluxetable}{cccc}
\tablecaption{New SN Photometry}\label{table_photometry}
\tablehead{\colhead{Name} & \colhead{Filter} & \colhead{JD-2450000} & \colhead{v}  \\ 
\colhead{} &\colhead{} & \colhead{(days)} & \colhead{(mag)}   } 

\startdata
SN~2008hs &  $v$   &    4805.05 &     17.42 $\pm$ 0.09 \\
SN~2008hs &  $b$   &    4805.05  &     17.27 $\pm$ 0.05 \\
SN~2008hs &  $u$   &    4805.04 &     16.80 $\pm$ 0.05 \\
SN~2008hs & uvw1 &    4805.04 &       19.69   $\pm$ 0.18 \\
SN~2008hs & uvm2 &    4805.05 &      20.89 $\pm$  0.30 \\
SN~2008hs & uvw2 &    4805.05  &    20.09   $\pm$    0.19 \\

\enddata
\tablecomments{The full table of photometry is available in the electronic version.}
\end{deluxetable}

%%%%%%%%%%%%%%%%%%%%%%%%%%%%%%%%%%%%%%%%%%%%%%%%%%% 

\begin{deluxetable}{llrllll}
\tablecaption{SN Parameters}\label{table_results}
\tablehead{\colhead{Name} &  \colhead{Distance\tablenotemark{1}} & \colhead{$E(B-V)$} & \colhead{\mb} & \colhead{$M_B$} & \colhead{Host} & \colhead{Morphology}   \\ 
\colhead{} & \colhead{Modulus}  & \colhead{Reddening}  & \colhead{ } & \colhead{  }   
& \colhead{Galaxy}    & \colhead{ } \\
\colhead{} & \colhead{(mag)} & \colhead{(mag)} & \colhead{(mag)} & \colhead{(mag)} & \colhead{ }   & \colhead{ }  } 

\startdata
SN2005cf & 32.59$ \pm $ 0.24 &  0.19$ \pm $ 0.04 &                   1.07 $ \pm $ 0.03 &  -19.84 $\pm$ 0.29 &   MCG-01-39-003 & S0 pec \\
SN2005ke & 31.70$ \pm $ 0.19 &  0.10$ \pm $ 0.03 &                    1.77 $ \pm $ 0.01 &  -17.21 $\pm$ 0.24 &  NGC 1371 & Sa \\
SN2006dd & 31.61$ \pm $ 0.08\tablenotemark{2} &  0.02$ \pm $ 0.05 &     1.08 $\pm$ 0.01   &   -19.43 $\pm$ 0.17  & NGC 1316 & E        \\
SN2007af & 32.31$ \pm $ 0.26 &  0.17$ \pm $ 0.03 &                   1.22 $ \pm $ 0.05 & -19.60 $\pm$ 0.30 &  NGC 5584 & Sc \\
SN2007cv & 33.07$ \pm $ 0.20\tablenotemark{3} &  0.21$ \pm $ 0.08 & 1.33 $ \pm $ 0.05 & -18.63 $\pm$ 0.40 &  IC 2597 & E \\
 SN2008Q & 31.74$ \pm $ 0.20\tablenotemark{4} &  0.10$ \pm $ 0.08 & 1.40 $ \pm $ 0.05 & -18.30 $\pm$ 0.40 &  NGC 524 & S02/Sa \\
SN2008ec & 34.16$ \pm $ 0.18 &  0.23$ \pm $ 0.08 &                    1.08 $ \pm $ 0.05 & -19.28 $\pm$ 0.39 &  NGC 7469 & Sab \\
SN2008hs & 34.26$ \pm $ 0.17 & -0.10$ \pm $ 0.09 &  2.0  $\pm$ 0.2  &  -18.19 $\pm$ 0.41  & NGC910   &   E  \\
SN2008hv & 33.76$ \pm $ 0.19 &  0.06$ \pm $ 0.08 &  1.2  $\pm$ 0.1  &  -18.94 $\pm$ 0.40  & NGC2765 & S0 \\
SNF080514\tablenotemark{5} & 35.02$ \pm $ 0.16 & -0.03$ \pm $ 0.08 & 1.2 $\pm$ 0.1 & -19.07 $\pm$ 0.39  & UGC8472 & S0 \\
SN2009ig & 32.73$ \pm $ 0.23 &  0.20$ \pm $ 0.08 & 0.85 $\pm$ 0.1  & -19.71 $\pm$ 0.53  & NGC1015 & SBa  \\
 SN2010Y & 34.64$ \pm $ 0.17 & -0.08$ \pm $ 0.15 & 1.70 $\pm$ 0.1   & -18.06 $\pm$ 0.66  & NGC3392 & E  \\
%
%\tablecomments{  }
\enddata
\tablenotetext{1}{Hubble flow distances are used except for those noted below.}
\tablenotetext{2}{Surface Brightness Fluctuations (SBF) distance from \citet{Blakeslee_etal_2009}.}
\tablenotetext{3}{SBF distance from \citet{Mieske_etal_2005}.}
\tablenotetext{4}{SBF distance from \citet{Jensen_etal_2003}.}
\tablenotetext{5}{SNF20080514-002}
\end{deluxetable}

%%%%%%%%%%%%%%%%%%%%%%%%%%%%%%%%%%%%%%%%%%%%%%%%%%% 

\begin{deluxetable}{lcccccc}

\tablecaption{Explosion Date and Companion Separation Limits}\label{table_results}
\tablehead{\colhead{Name} &   \colhead{Explosion} & \colhead{Epoch of first} & \colhead{a$_{13}=0.2$} & \colhead{a$_{13}=0.2$}  & \colhead{a$_{13}=2$} & \colhead{a$_{13}=2$} \\ 
\colhead{}   & \colhead{Date} & \colhead{UV observation }   
& \colhead{50\% limit on $\theta$}    & \colhead{95\% limit on $\theta$} 
& \colhead{50\%  limit }    & \colhead{95\% limit } \\
\colhead{}  & \colhead{(days)} & \colhead{(days)} & \colhead{(degrees)}   & \colhead{(degrees)} & \colhead{(degrees)}   & \colhead{(degrees)} } 

\startdata
SN2005cf & 3517.31 $^{+0.06} _{-0.06}$ &  8.24 &    $>$ 0.0 &    $>$ 0.0 &  $>$164.2 &  $>$158.5 \\
SN2005ke & 3685.77 $^{+0.24} _{-0.09}$ &  3.49 &  $>$165.9 &  $>$161.8 &  $>$177.4 &  $>$176.2 \\
SN2006dd & 3902.70 $^{+0.02} _{-0.14}$ &  4.72 &   $>$ 85.7 &   $>$37.0 &  $>$169.4 &  $>$165.2 \\
SN2007af & 4157.12 $^{+0.58} _{-0.07}$ &  5.23 &  $>$101.0 &   $>$43.0 &  $>$171.2 &  $>$167.5 \\
SN2007cv & 4276.63 $^{+0.50} _{-0.31}$ &  5.64 &    $>$ 7.0 &    $>$0.0 &  $>$165.5 &  $>$158.0 \\
 SN2008Q & 4491.57 $^{+0.41} _{-0.24}$ &  5.06 &  $>$103.3 &   $>$24.3 &  $>$171.3 &  $>$165.1 \\
SN2008ec & 4657.93 $^{+0.49} _{-0.32}$ &  5.26 &    $>$ 0.0 &    $>$0.0 &  $>$157.8 &  $>$146.1 \\
SN2008hs & 4799.23 $^{+0.17} _{-0.40}$ &  4.40 &  $>$147.8 &  $>$113.6 &  $>$175.4 &  $>$171.7 \\
SN2008hv & 4801.82 $^{+0.24} _{-0.09}$ &  3.37 &  $>$155.1 &  $>$134.8 &  $>$175.3 &  $>$172.1 \\
SNF080514\tablenotemark{1} & 4598.95 $^{+0.26} _{-1.88}$ &  7.53 &    $>$ 0.0 &    $>$ 0.0 &  $>$145.1 &  $>$113.8 \\
SN2009ig & 5063.31 $^{+0.10} _{-0.05}$ &  2.03 &  $>$154.7 &  $>$138.0 &  $>$174.0 &  $>$170.4 \\
 SN2010Y & 5232.62 $^{+0.34} _{-0.40}$ &  4.88 &   $>$98.9 &    $>$ 0.0 &  $>$171.0 &  $>$155.2 \\
%
%\tablecomments{  }
\enddata
\tablenotetext{1}{SNF20080514-002}
\end{deluxetable}

%%%%%%%%%%%%%%%%%%%%%%%%%%%%%%%%%%%%%%%%%%%%%%%%%%%%%%%
%  %%%%%%%     %%         %%%%%% %%%%%%%    %%%%%                 
%  %%    %%    %%        %%    %%  %%      %%
%  %%    %%    %%       %%      %% %%      %%   
%  %%%%%%%     %%       %%      %% %%       %%
%  %%          %%       %%      %% %%         %%              
%  %%          %%        %%    %%  %%        %%
%  %%          %%%%%%%    %%%%%%   %%    %%%% 
%%%%%%%%%%%%%%%%%%%%%%%%%%%%%%%%%%%%%%%%%%%%%%%%%%%%%%%%%%%

%\clearpage 
%\clearpage 

%%%%%%%%%%%%%%%%%%%%%%%%%%%%%%%%%%%%%%%%%%%%%%%%%%%%%%%%%%%%%%%%%%%%%%%%%%%%%%%%% 

%\clearpage 
\begin{figure} 
%\resizebox{16cm}{!}{\rotatebox{90}{\includegraphics*{f2.eps}  } }
\plottwo{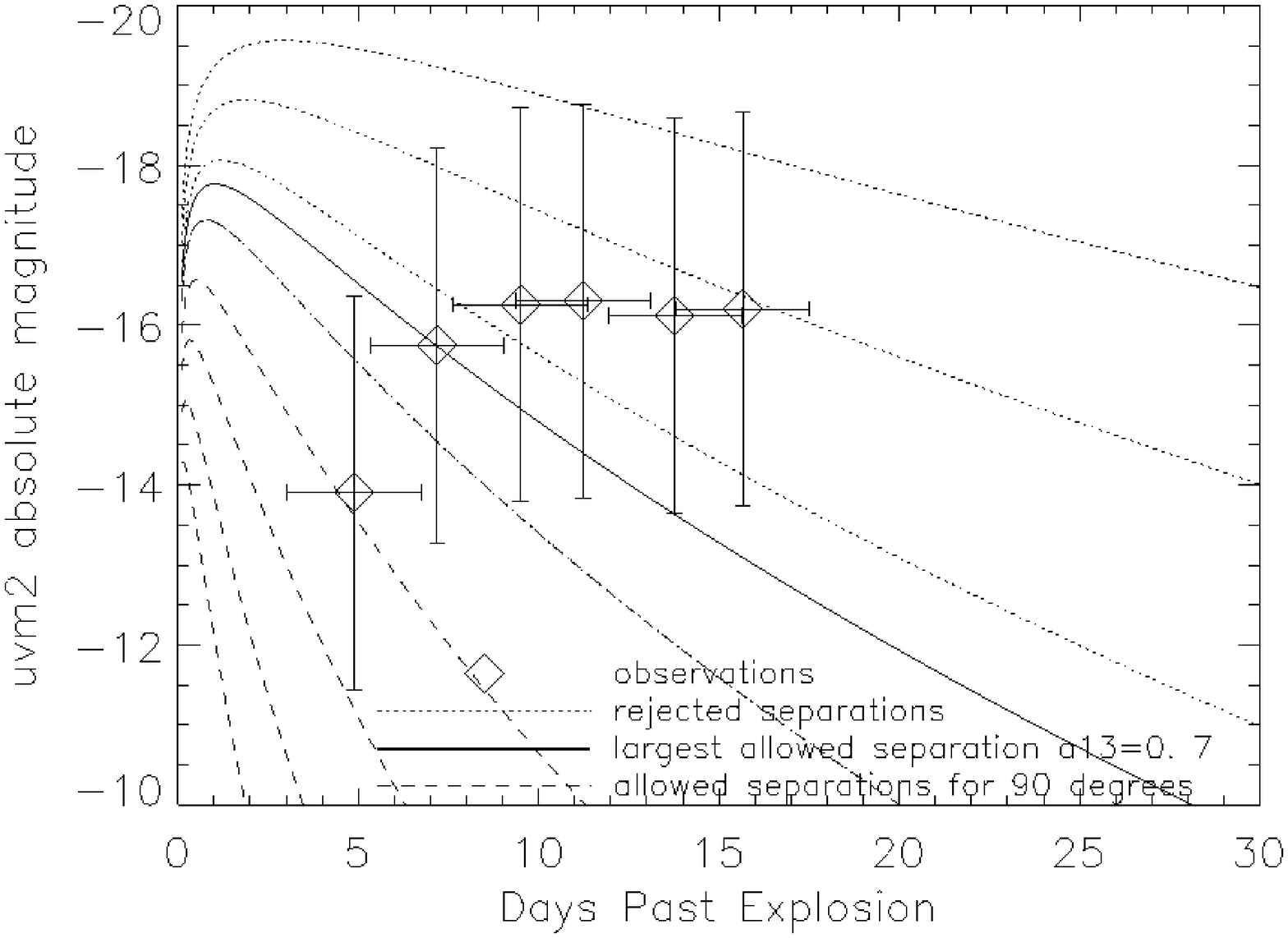}{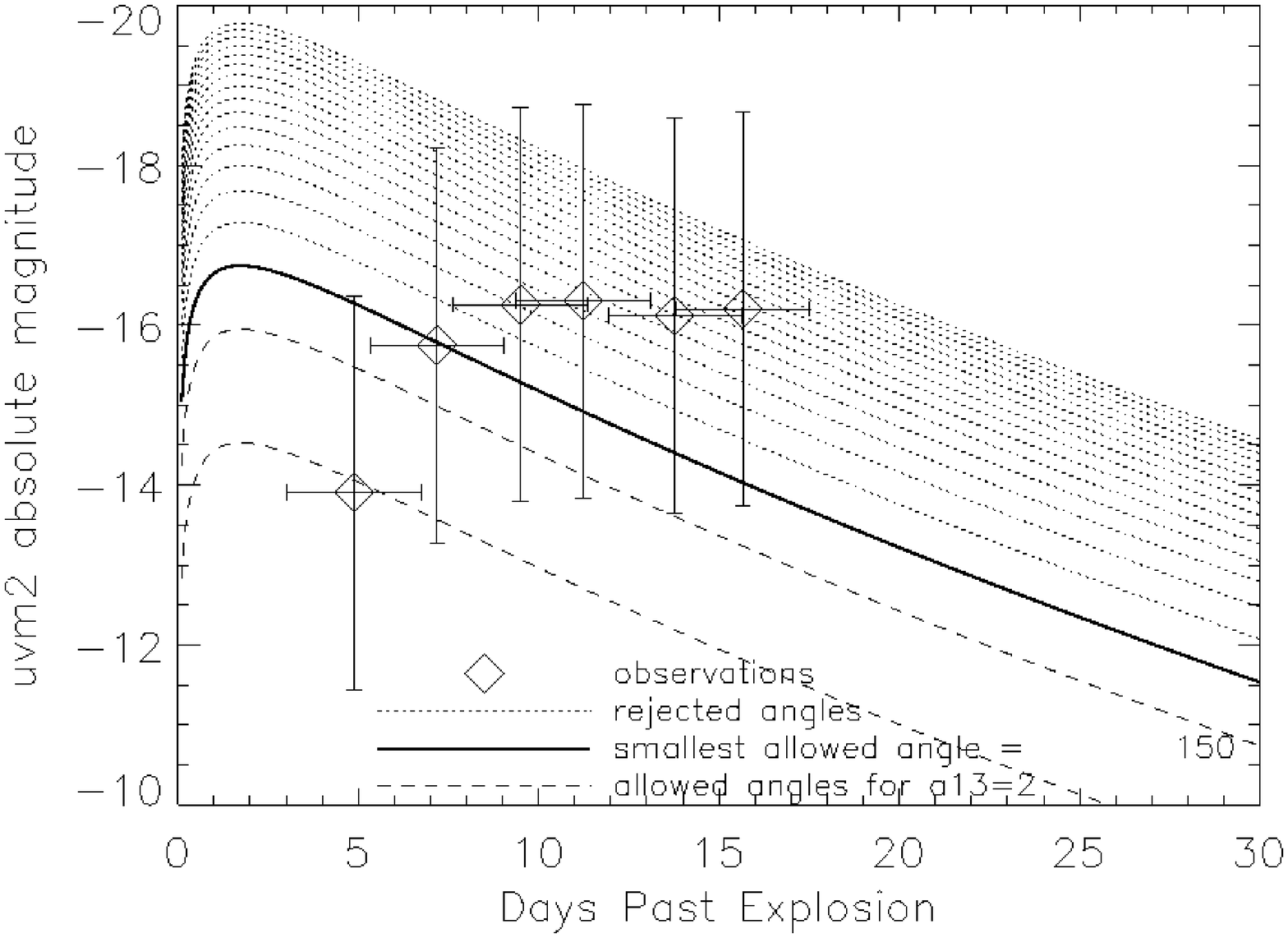}
\caption[Results]
        {{\it Left:} A series of models with separation distances a13=0.01 to 6 (spaced geometrically) 
viewed at 90 degrees compared to the uvm2 absolute magnitudes of SN~2010Y.  The error bars displayed correspond to 95\% confidence limits on the explosion date (x errors) and the photometric, extinction, and distance modulus in quadrature (y errors).
The models shown with dashed lines are allowed in 95\% of the realizations (for this viewing angle), while those shown with dotted lines 
are rejected because they are brighter than at least 5\% of the realizations.  The brightest curve allowed in 95\% of the realizations, 
corresponding to the largest allowed separation distance a$_{13}$=0.7, is shown with a solid line.
{\it Right:} The a$_{13}$=2 RG model from different viewing angles (0 to 170 degrees, spaced by 10 degrees) compared to the observations of SN~2010Y.  The angles (for this separation distance) shown with dashed lines are allowed in 95\% of the realizations, 
while those shown with dotted lines are rejected because they are brighter than 5\% of the realizations. The brightest allowed curve, corresponding to 150 degrees, is shown with a solid line.
 } \label{plot_modelsangles}    
\end{figure} 

 %\clearpage 
\begin{figure} 
%\resizebox{16cm}{!}{\rotatebox{90}{\includegraphics*{f2.eps}  } }
\plottwo{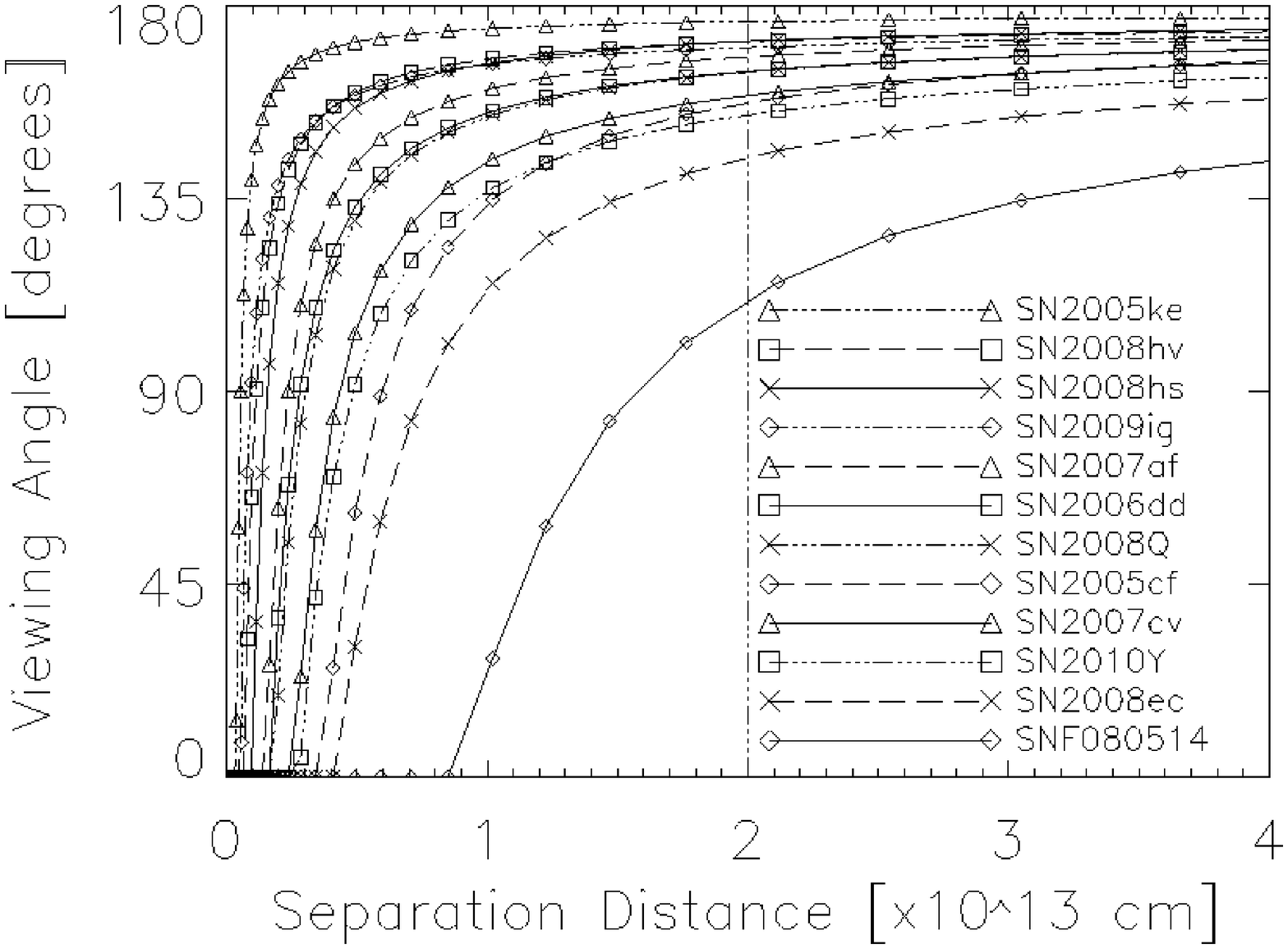}{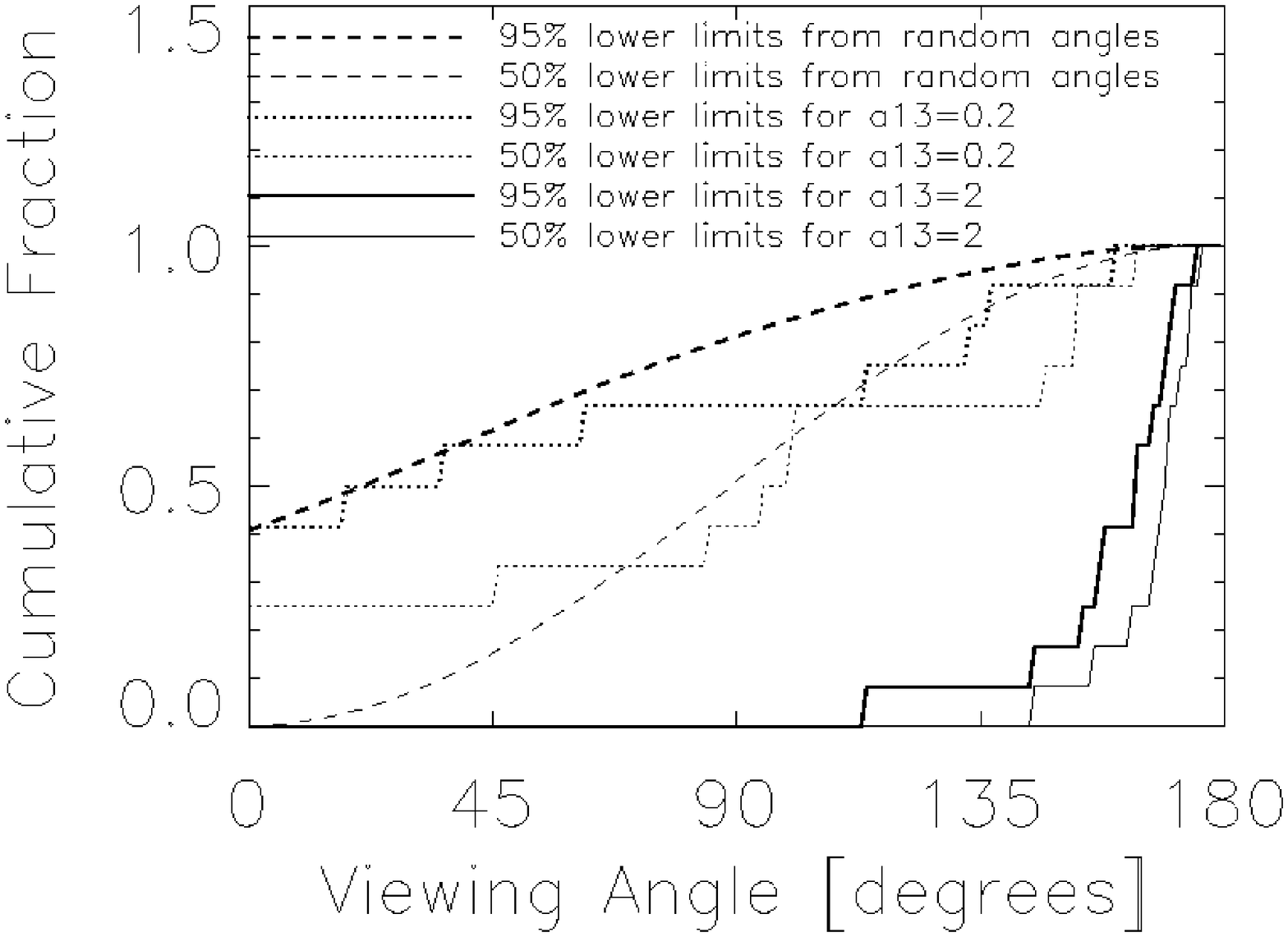} 
\caption[Results]
        {{\it Left:} Separation distance-viewing angle constraints for our sample of SNe.  
These constraints are the strictest from all the filters and epochs considered individually.
The regions under the curves are excluded by the observations at 95\% confidence.  
The vertical dashed line at a$_{13}$=2 corresponds to a 1~M$_\sun$ RG.  
{\it Right:} Cumulative distribution functions of the 95\% lower limit on the 
viewing angle for models with separation distances 
of 0.2 and $2 \times 10^{13} $  cm.  Also shown are the cumulative distribution 
functions expected from random viewing angles for the respective models 
subjected to the same uncertainties as the data. 

} \label{plot_sne_mc}
\end{figure} 
%\clearpage 

 %\clearpage 
\begin{figure} 
%\resizebox{16cm}{!}{\rotatebox{90}{\includegraphics*{f2.eps}  } }
\plottwo{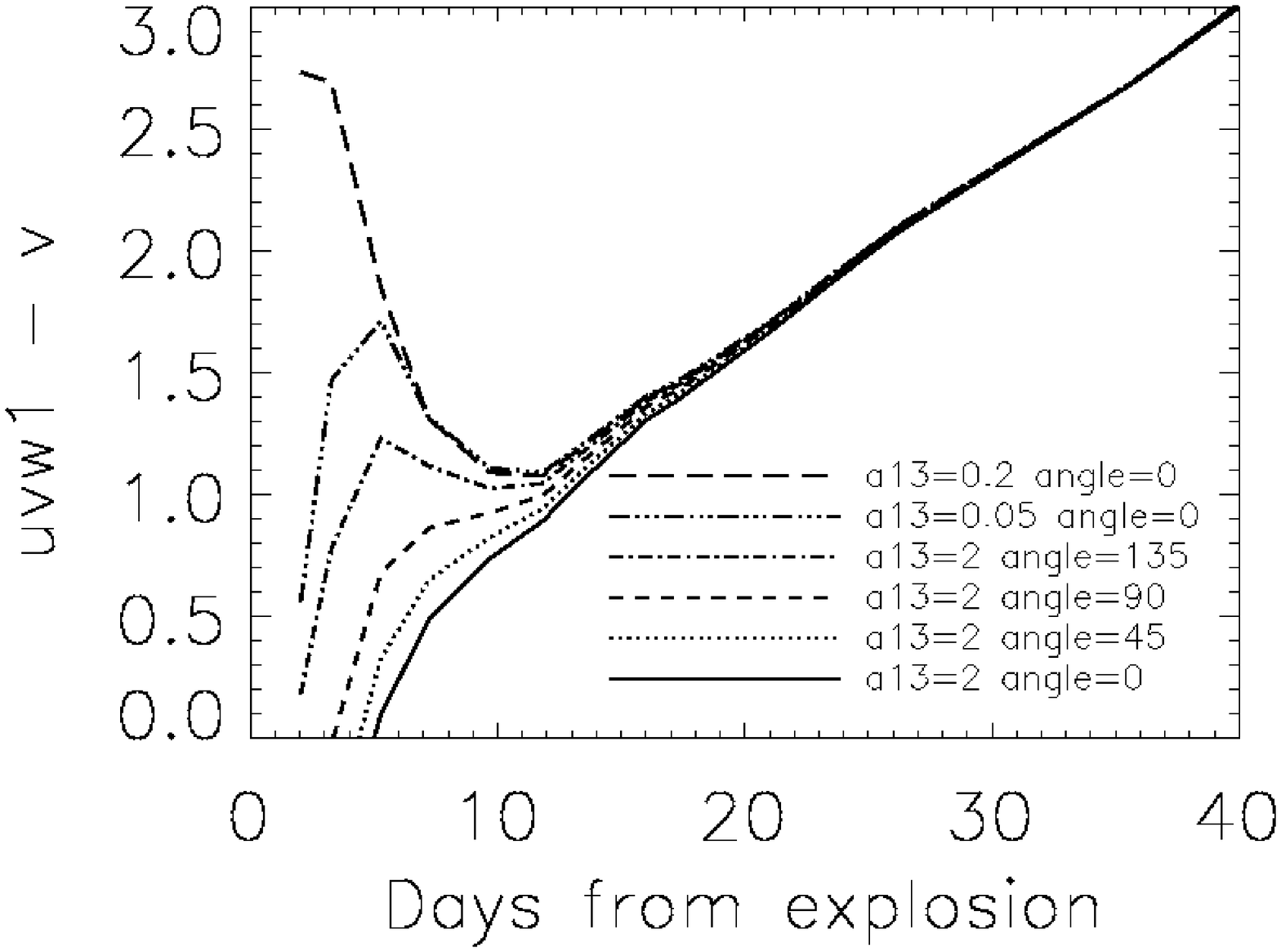}{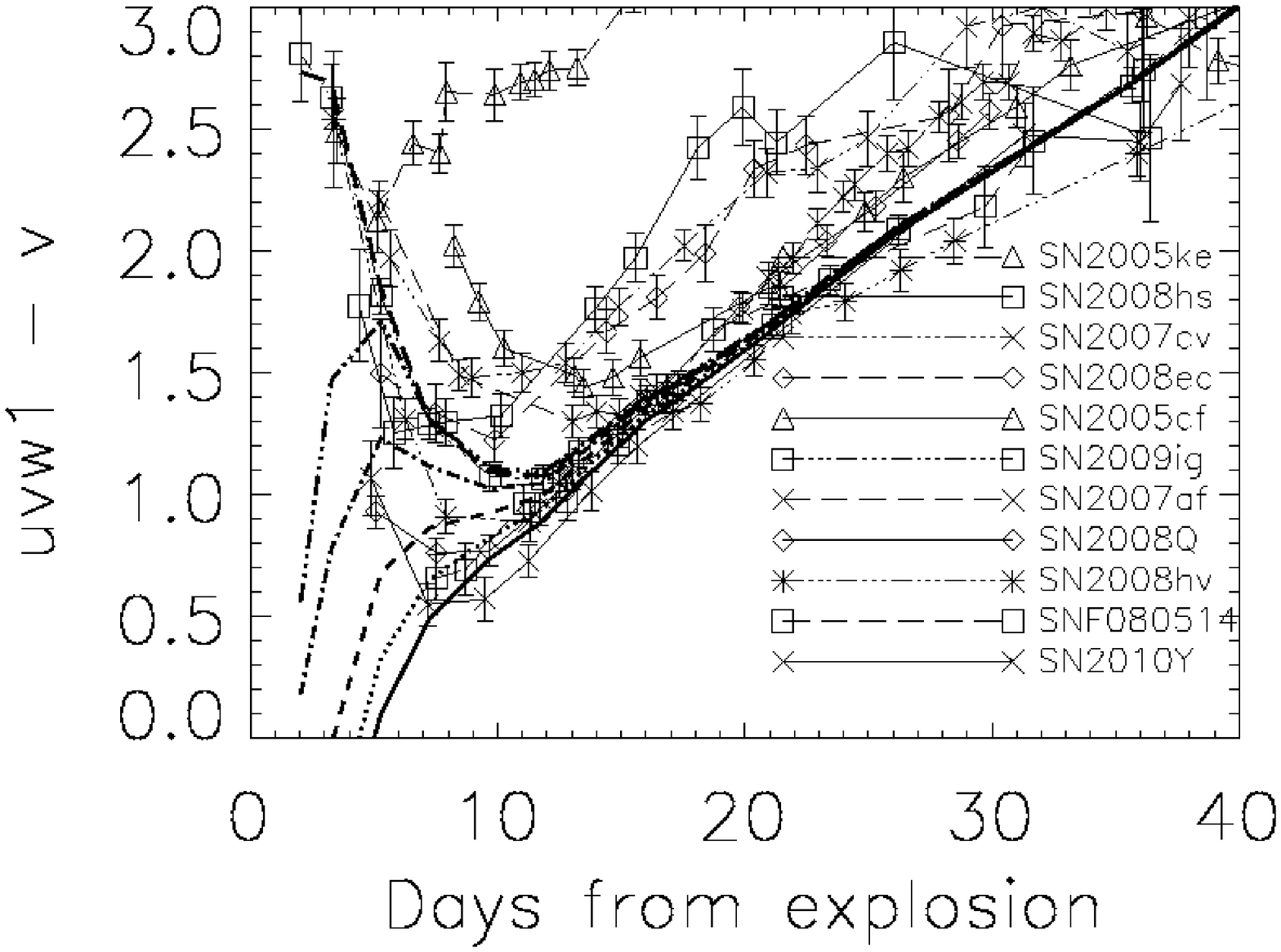} 
\caption[Results]
        {{\it Left:} uvw1-v colors of various shock models added to the 
observations of SN~2009ig (whose observations began the soonest after explosion).  
{\it Right:} Observed uvw1-v colors of our SN sample compared to the models shown to the left.

} \label{plot_colors}
\end{figure} 
%\clearpage 

\end{document}